\documentclass[pdflatex,sn-nature,iicol]{sn-jnl}

\usepackage{graphicx}

\usepackage{multirow}
\usepackage{amsmath,amssymb,amsfonts}
\usepackage{anyfontsize}
\usepackage{mathrsfs}
\usepackage[title]{appendix}
\usepackage{xcolor}
\usepackage{textcomp}
\usepackage{booktabs}
\usepackage{algorithm}
\usepackage{algorithmicx}
\usepackage{algpseudocode}
\usepackage{listings}
\usepackage{float}
\usepackage{stfloats}  
\usepackage{afterpage} 

\usepackage{geometry}
\makeatletter
\g@addto@macro\maketitle{\restoregeometry}
\makeatother

\usepackage{fancyhdr}

\fancypagestyle{plain}{%
    \fancyhf{}%
    \fancyfoot[C]{\thepage}
}

\fancypagestyle{fancy}{%
    \fancyhf{}%
    \fancyfoot[C]{\thepage}
}

\makeatletter
\let\ps@pprintTitle\ps@plain
\makeatother

\theoremstyle{thmstyleone}

\theoremstyle{thmstyletwo}

\theoremstyle{thmstylethree}

\begin{document}

\title[Short Title]{ Landau Zener Interaction Enhanced Quantum Sensing in Spin Defects of Hexagonal Boron Nitride}

\author*[1]{\fnm{Mohammad Abdullah} \sur{Sadi}}\email{msadi@purdue.edu}
\author[2]{\fnm{Tiamike} \sur{Dudley}}
\author[3]{\fnm{Luca} \sur{Basso}}
\author[4]{\fnm{Thomas} \sur{Poirier}}
\author[4]{\fnm{James H.} \sur{Edgar}}
\author[3]{\fnm{Jacob} \sur{Henshaw}}
\author[1]{\fnm{Peter A.} \sur{Bermel}}
\author*[1,5,6,7]{\fnm{Yong P.} \sur{Chen}}\email{yongchen@purdue.edu}
\author*[3]{\fnm{Andrew} \sur{Mounce}}\email{amounce@sandia.gov}

\affil[1]{\orgdiv{Elmore Family School of Electrical and Computer Engineering}, \orgname{Purdue University}, 
\orgaddress{\city{West Lafayette}, \postcode{47907}, 
\state{Indiana}, \country{USA}}}

\affil[2]{\orgdiv{Electrical and Computer Engineering Department}, \orgname{The University of New Mexico}, 
\orgaddress{\city{Albuquerque}, \postcode{87106}, 
\state{New Mexico}, \country{USA}}}

\affil[3]{\orgdiv{Center for Integrated Nanotechnologies}, \orgname{Sandia National Laboratories}, 
\orgaddress{\city{Albuquerque}, \postcode{87123}, 
\state{New Mexico}, \country{USA}}}

\affil[4]{\orgdiv{Tim Taylor Department of Chemical Engineering}, \orgname{Kansas State University}, 
\orgaddress{\city{Manhattan}, \postcode{66506}, 
\state{Kansas}, \country{USA}}}

\affil[5]{\orgdiv{Department of Physics and Astronomy}, \orgname{Purdue University}, 
\orgaddress{\city{West Lafayette}, \postcode{47907}, 
\state{Indiana}, \country{USA}}}

\affil[6]{\orgdiv{Institute of Physics and Astronomy and Villum Center for Hybrid Quantum Materials and Devices}, \orgname{Aarhus University}, 
\orgaddress{\city{Aarhus}, \postcode{8000}, \country{Denmark}}}

\affil[7]{\orgdiv{WPI-AIMR International Research Center for Materials Sciences}, \orgname{Tohoku University}, 
\orgaddress{\city{Sendai}, \postcode{980-8577}, \country{Japan}}}

\abstract{Negatively charged boron vacancies (V$_{\text{B}}^{-}$) in hexagonal boron nitride (hBN) comprise a promising quantum sensing platform, optically addressable at room temperature and transferrable onto samples. However, broad hyperfine-split spin transitions of the ensemble pose challenges for quantum sensing with conventional resonant excitation due to limited spectral coverage. While isotopically enriched hBN using $^{10}$B and $^{15}$N isotopes (h$^{10}$B$^{15}$N) exhibits sharper spectral features, significant inhomogeneous broadening persists. We demonstrate that, implemented via frequency modulation on an FPGA, a frequency-ramped microwave pulse achieves around 4-fold greater $|0\rangle\rightarrow|-1\rangle$ spin-state population transfer and thus contrast than resonant microwave excitation and thus 16-fold shorter measurement time for spin relaxation based quantum sensing. Quantum dynamics simulations reveal that an effective two-state Landau-Zener model captures the complex relationship between population inversion and pulse length with relaxations incorporated. Our approach is robust and valuable for quantum relaxometry with spin defects in hBN in noisy environments.}

\maketitle


\section*{Introduction}\label{sec:intro}

\begin{figure*}[!b]
\centering
\includegraphics[width=0.98\textwidth]{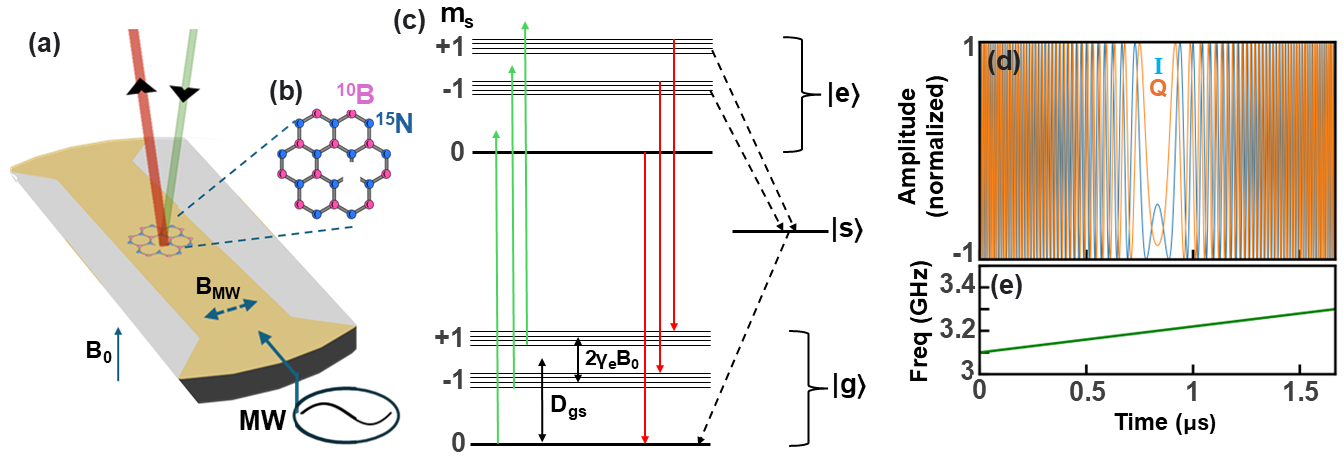}
\caption{Experimental configuration, energy levels and mechanism of generating frequency ramp to be applied to V$_\text{B}^-$ defects in isotopically enriched hexagonal boron nitride (h$^{10}$B$^{15}$N). (a) Gold on sapphire device generating linearly polarized microwaves. Center: multilayer h$^{10}$B$^{15}$N containing spin defects. (b) Lattice structure of h$^{10}$B$^{15}$N with V$_\text{B}^-$. (c) Electronic energy level diagram of V$_\text{B}^-$ showing S=1 triplet states with four hyperfine levels arising from total 3/2 spin of the three $^{15}$N  surrounding the defect, for non-zero m$_\text{s}$ spin states, and S=0 metastable state mediating ODMR (green: optical excitation, red: radiative decay, dashed black: intersystem crossing decay). $|e\rangle$, $|s\rangle$ and $|g\rangle$ represent excited, metastable, and ground states. (d) I and Q values fed into FPGA is convoluted with the carrier frequency to generate the frequency ramp (e) of $\Delta f = 200$ MHz centered at $f_0 = 3.2$ GHz.}
	\label{fig:figure1}
\label{fig:figure1}
\end{figure*}

Quantum sensing with optically addressable spin defects in solids has enabled remarkable advances in nanoscale magnetometry, with applications spanning materials science, condensed matter physics, and biology\cite{degen2017quantum,barry2020sensitivity,schirhagl2014nitrogen}. The nitrogen-vacancy (NV) center in diamond has served as the leading platform for solid-state quantum sensing, combining remarkably long coherence times at room temperature with single-spin magnetic resonance at atomic-scale resolution\cite{schirhagl2014nitrogen,rondin2014magnetometry}. Yet diamond's bulk crystal structure imposes limitations for practical sensing applications. Fabricating near-surface NV centers at targeted locations requires requires shallow ion implantation or delta-doping, which significantly reduces coherence times due to charge noise from surface defects\cite{romach2015spectroscopy,sangtawesin2019origins}, while diamond's rigidity limits conformal contact with non-planar samples.

Two-dimensional van der Waals materials overcome these constraints while enabling mechanical transfer onto arbitrary substrates\cite{azzam2021prospects,caldwell2019photonics}. Among 2D materials, hexagonal boron nitride (hBN) has emerged as a particularly promising quantum sensing platform due to its wide bandgap, exceptional chemical inertness, thermal stability, and optical transparency\cite{azzam2021prospects,cassabois2016hexagonal,kianinia2017robust}. The negatively charged boron vacancy (V$_{\text{B}}^{-}$) in hBN, identified in 2020 as an optically addressable spin-1 defect\cite{gottscholl2020initialization}, exhibits room-temperature quantum coherence with optically detected magnetic resonance (ODMR)\cite{gao2021high,yu2022excited,mathur2022excited}, achieving shorter coherence times than NV diamond but enabling atomic-scale proximity between the sensor and sample\cite{ivady2020ab,reimers2020photoluminescence,chen2021photophysical}.

Strong hyperfine interactions between the V$_{\text{B}}^{-}$ electron spin and surrounding nuclear spins in natural hBN severely broaden spin resonances, with multiple hyperfine transitions producing overlapping features\cite{gottscholl2020initialization,reimers2020photoluminescence}. For ODMR spectroscopy, spin-state-selective excitation using circularly polarized microwaves has demonstrated improved contrast at low and zero magnetic fields\cite{sadi2025spinstate}. Isotopic engineering with h$^{10}$B$^{15}$N has demonstrated substantial linewidth reductions\cite{mu2025magnetic,fraunie2025charge}, enabling applications including magnetic imaging under high pressure\cite{mu2025magnetic} and charge state control in van der Waals heterostructures\cite{fraunie2025charge}. While isotopic purification substantially reduces linewidths compared to natural hBN, remaining hyperfine structure still produces significant spectral broadening, resulting in linewidths much broader than the sub-MHz linewidths from NV centers in isotopically purified diamond\cite{balasubramanian2009ultralong, Kleinsasserlifetime}. For coherent spin manipulation, narrow-band resonant pulses address only a small fraction of such inhomogeneously broadened spectrum at any given frequency, constraining population transfer efficiency across the full spectral distribution.

Landau-Zener (LZ) transitions provide a fundamentally different approach to quantum state manipulation in spectrally complex systems\cite{landau1932,zener1932,shevchenko2010review,ivakhnenko2023review}. The theory, independently discovered in 1932\cite{landau1932,zener1932,stuckelberg1932,majorana1932}, describes nonadiabatic population transfer when a system is driven through an avoided level crossing, with swept excitation driving transitions across a range of frequencies to naturally accommodate inhomogeneous broadening. Nuclear magnetic resonance provided early validation through adiabatic passage techniques\cite{garwood2001frequency}, achieving population transfer with fidelity limited by relaxation rather than excitation bandwidth. Ultracold atomic systems have since exploited LZ transitions with tunable parameter control\cite{olson2014tunable,rogora2024zeromagnetic,qian2013manybody}. Nitrogen-vacancy centers in diamond have particularly benefited from LZ-based control: Fuchs et al. achieved high-fidelity room-temperature quantum memory using rapid LZ transitions at hyperfine-mediated avoided crossings\cite{fuchs2011quantum}, Kavtanyuk et al. demonstrated substantial $^{13}$C nuclear spin hyperpolarization through microwave-driven LZ transitions with optimum sweep widths spanning the full hyperfine manifold\cite{kavtanyuk2025hyperpolarization}, and Henshaw et al. developed magnetic field sweeps for robust polarization control\cite{henshaw2019dnp}. Semiconductor quantum dots have similarly leveraged LZ dynamics for ultrafast control\cite{cao2013ultrafast,petta2010beam}.

In this work, inspired by successful implementations across diverse quantum platforms, we apply Landau-Zener transitions to V$_{\text{B}}^{-}$ defects in isotopically substituted h$^{10}$B$^{15}$N, where frequency sweeps can traverse a wider spectral range across the hyperfine-broadened spectrum compared to fixed-frequency resonant pulses. We implement frequency-ramped microwave using an FPGA and deliver it to the spin defects through a gold-on-sapphire device (Fig.~\ref{fig:figure1}(a)-(b)). We achieve four-fold greater population transfer compared to optimized resonant excitation, translating to more than 16-fold reduction in measurement time for quantum sensing. We simulate the sweep dynamics using density matrix evolution in QuTiP\cite{johansson2013qutip}, incorporating relaxations and effective coupling strength, which closely reproduces the experimental results and reveals that despite the complex hyperfine manifold structure, the system is well-described by an effective two-level model. Our approach enhances applicability of V$_{\text{B}}^{-}$ defects in hBN for quantum sensing in scenarios where high laser intensities or long averaging times are impractical, such as characterization of photosensitive samples or biological specimens in solution.

\section*{Results and Discussion}\label{sec:results}
\subsection*{Enhanced Population Transfer with Frequency Sweep}
\begin{figure*}[!t]
\centering
\includegraphics[width=0.98\textwidth]{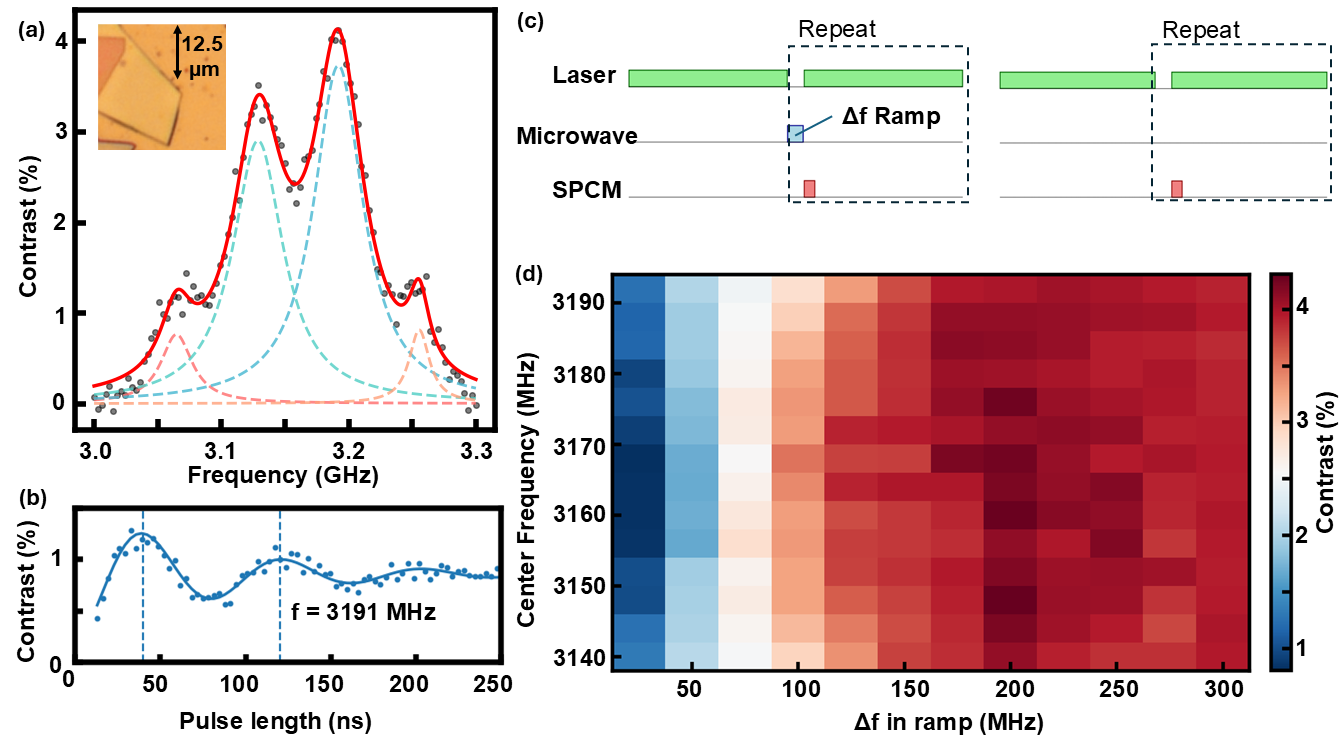}
\caption{ODMR, Rabi oscillation, and frequency-ramped microwave pulses for enhanced spin inversion (population transfer) across broad hyperfine-split transitions in h$^{10}$B$^{15}$N. (a) Continuous wave ODMR spectrum showing the $|0\rangle$ to $|-1\rangle$ transition isolated by applied magnetic field, with solid line being composite of dashed hyperfine Lorentzian peaks (inset: optical image of the hBN flake). (b) Rabi oscillation at resonant frequency of 3191 MHz at the third hyperfine peak, showing maximum contrast of approximately 1.2\% for $\pi$-pulse. (c) Pulse sequence diagram illustrating the ramped frequency microwave pulse application scheme, with contrast measured by comparison to the pulse sequence without microwave. (d) Frequency ramp demonstrating contrast enhancement beyond 1.2\%, robust to variation in ramp center frequency, with optimal performance at $\Delta f \approx 200$ MHz. Rabi oscillation and frequency ramp measurements (b, d) performed at MW gain = 25000 and collection time = 1 $\mu$s, and ODMR spectrum in (a) also obtained at the same MW gain. The microwave ramp duration used in (c, d) is T=1.67 $\mu$s.}
\label{fig:figure2}
\end{figure*}

The spin dynamics of V$_\text{B}^-$ centers in h$^{10}$B$^{15}$N (Fig. \ref{fig:figure1}(b)) are described by a ground-state Hamiltonian $H_{gs}$ containing Zero Field Splitting (ZFS), electron Zeeman, and hyperfine coupling terms \cite{cluaprovost2023isotopic,gong2024isotope}:
\begin{equation}
\begin{split}
H_{gs} = D_{gs} \left[ S_z^2 - \frac{S(S+1)}{3} \right] + E_{gs} \left( S_x^2 - S_y^2 \right) & \\ + \gamma_e \mathbf{B} \cdot \mathbf{S} + \sum_{k=1,2,3} \mathbf{S}\cdot\bar{\bar{A}}_k\cdot\mathbf{I}_k
\end{split}
\end{equation}
where $D_{gs} \approx h \times 3.48 \text{ GHz}$ represents the longitudinal ZFS and $E_{gs} \approx h \times 50 \text{ MHz}$ the transverse ZFS. The electron spin operators $S_z$, $S_x$, and $S_y$ correspond to a spin-1 system ($S=1$), while $\gamma_e = g \frac{e}{2m_e\hbar}$ denotes the electron gyromagnetic ratio with Landé $g$-factor $g = 2$. In isotopically enriched h$^{10}$B$^{15}$N, the three nearest-neighbor nitrogen atoms contribute nuclear spin-1/2 operators $\mathbf{I}_k$ \cite{cluaprovost2023isotopic,sasaki2023nitrogen}. The three spin-1/2 $^{15}$N nuclei surrounding the vacancy produce four resolved hyperfine transitions spaced by $A_k \approx h \times 64 \text{ MHz}$ (Fig. \ref{fig:figure1}(c) and Fig. \ref{fig:figure2}(a)) \cite{sasaki2023nitrogen,haykal2022decoherence}. The presence of $^{10}$B reduces spectral linewidths compared to $^{11}$B (more abundant in natural hBN) due to the lower nuclear gyromagnetic ratio of $^{10}$B ($\gamma_n(^{10}\text{B})/2\pi \approx 4.58$ MHz/T) compared to $^{11}$B ($\gamma_n(^{11}\text{B})/2\pi \approx 13.66$ MHz/T), despite $^{10}$B having a higher nuclear spin ($I = 3$ vs $I = 3/2$) \cite{cluaprovost2023isotopic,haykal2022decoherence}.

Under a perpendicular magnetic field $B_0$, the spin transition frequencies between $m_s = 0$ and $m_s = \pm1$ levels follow:
\begin{equation}
	f_{\pm} = D_{gs}/h \pm \sqrt{E_{gs}^2 + (\gamma_e B_0)^2}/h
\end{equation}

Microwave frequency ramps are generated using QICK-DAWG\cite{riendeau2023quantum}, an FPGA-based quantum instrumentation control kit. To generate a baseband signal with quadratic phase:
\begin{equation}
	\phi(t) = -\tfrac{\Delta\omega}{2}t + \tfrac{\Delta\omega}{2T}t^2
\end{equation}
where $\Delta\omega = 2\pi\Delta f$ and $\Delta f$ is the chirp bandwidth, the I/Q waveforms of duration $T$ are synthesized as (Fig. \ref{fig:figure1}(d)):
\begin{equation}
	I(t) = \cos\phi(t), \qquad Q(t) = -\sin\phi(t)
\end{equation}
These encode the complex baseband signal:
\begin{equation}
	I(t) + jQ(t) = e^{-j\phi(t)}
\end{equation}
The instantaneous frequency, related to phase by $f(t) = \frac{1}{2\pi}\frac{d\phi}{dt}$, gives the baseband sweep:
\begin{equation}
	f(t) = -\frac{\Delta f}{2} + \frac{\Delta f}{T}t
\end{equation}
Upconversion to carrier frequency $f_0$ shifts the spectrum, yielding (Fig. \ref{fig:figure1}(e)):
\begin{equation}
	f_{\text{output}}(t) = f_0 + f(t) = f_0 - \frac{\Delta f}{2} + \frac{\Delta f}{T}t
\end{equation}

As shown in Fig.~\ref{fig:figure2}(a), a perpendicular magnetic field $B_0$ is applied such that the center of the $|0\rangle$ to $|-1\rangle$ transition is positioned at $3160~\text{MHz}$. The \text{ODMR} spectrum reveals the characteristic hyperfine structure with four resolved transitions of the $\sim60~\text{nm}$ thick $\text{h}\,^{10}\text{B}\,^{15}\text{N}$ flake shown in the inset of the figure, and all subsequent measurements are performed on the flake with green laser irradiation intensity of $8.5~\text{mW}/\mu\text{m}^2$ at room temperature.

To characterize the traditional resonant excitation driven inversion, we perform Rabi oscillation measurements at the resonant frequency of 3191 MHz corresponding to the third hyperfine peak (Fig. \ref{fig:figure2}(b)). At a microwave gain of 25000, we observe coherent oscillations with a $\pi$-pulse duration of approximately 40 ns. The experimental sequence begins with optical initialization: a 15 $\mu$s green laser pulse populates the $|0\rangle$ ground state. Following the microwave manipulation pulse, we apply green laser illumination while simultaneously collecting red photoluminescence using a single photon counter module (SPCM) for a collection time of 1 $\mu$s. The contrast, defined as the normalized difference between microwave-off and microwave-on signals ($\text{Contrast} = \frac{I_{\text{off}} - I_{\text{on}}}{I_{\text{off}}}$) averaged over collection time, reaches approximately 1.2\% for the resonant $\pi$-pulse under these conditions. For the $\pi$-pulse length of 40 ns, it covers approximately 25 MHz spectral bandwidth. The $\pi$-pulse selectively thus addresses only a single hyperfine transition, while the other transitions, spaced 64 MHz apart, remain minimally unaffected.

To achieve broader spectral coverage and more robust spin manipulation, we replace the resonant $\pi$-pulse with frequency-ramped microwave pulses (Fig. \ref{fig:figure2}(c)). We vary the ramp center frequency f$_\text{0}$ from 3140 to 3190 MHz and the chirp bandwidth $\Delta f$ from 25 to 300 MHz, while maintaining a fixed ramp duration $T = 1.67$ $\mu$s and microwave gain of 25000. 

The probability for Landau-Zener adiabatic transitions is approximated by \cite{zener1932,landau1932,shevchenko2010review}:
\begin{equation}
	P_{\text{LZ}} = 1 - \exp\left(-\frac{2\pi\Omega^2}{\alpha}\right)
\end{equation}
where $\Omega$ is the Rabi frequency and $\alpha = d\omega/dt$ is the frequency sweep rate. This equation assumes that relaxation can be neglected. For a fixed sweep time $T$, the sweep rate increases with chirp bandwidth ($\alpha = \Delta\omega/T = 2\pi\Delta f/T$), which can compromise adiabaticity. Thus, a larger $\Delta f$ implies a faster sweep that may reduce the adiabatic transition probability. However, increasing $\Delta f$ results in broader spectral coverage and thus enables simultaneous addressing of multiple hyperfine transitions. At the current optimal parameters ($\Delta f \approx 200$ MHz centered at 3160 MHz), the first and fourth hyperfine transitions lie near the edges of the frequency sweep. These edge transitions experience minimal detuning during most of the ramp and consequently undergo limited inversion. Further increasing $\Delta f$ would extend the sweep range to better encompass all four hyperfine lines, though at the cost of reduced adiabaticity. Additionally, spin relaxation times impose another constraint on the optimal $\Delta f$: longer ramp durations required for larger $\Delta f$ at fixed adiabaticity may allow significant relaxation to occur during the sweep, degrading the overall transfer efficiency, as discussed later in the manuscript.

The color map in Fig. \ref{fig:figure2}(d) reveals the optimization landscape for these competing factors. We observe a maximum contrast of 4.5\% at a ramp center frequency of 3160 MHz with $\Delta f = 200$ MHz, representing a 3.7-fold enhancement over the resonant $\pi$-pulse contrast. Quantitative prediction of the optimal contrast from first principles is challenging due to the complex interplay between the four-transition hyperfine structure, inhomogeneous broadening, finite T$_1$ and T$_2$ relaxation during the sweep, and competing requirements for adiabaticity versus relaxation mitigation, requiring careful experimental navigation to identify the optimal parameters. Remarkably, the frequency-ramped approach exhibits substantially improved robustness: while the resonant $\pi$-pulse shows strong sensitivity to the drive frequency, the ramped pulse maintains similar inversion efficiency across the entire range of tested center frequencies. This robustness stems from the adiabatic nature of the frequency sweep, which ensures effective population transfer even when the instantaneous frequency is detuned from exact resonance. The broad region of high contrast in Fig. \ref{fig:figure2}(d) demonstrates that the frequency ramped pulse  provides enhanced performance and operational stability under realistic experimental conditions with finite microwave power and fixed ramp duration.

\begin{table}[h]
	\centering
	
	\caption{$\pi$-pulse characterization with microwave gain}
	\label{tab:your_label}
	\begin{tabular}{cccc}
		\toprule
		MW Gain & $\pi$-pulse length & $\Omega_{\text{nominal}}/2\pi$ & Contrast \\
		\midrule
		30000 & 38 ns & 13.2 MHz & 1.28\% \\
		25000 & 40 ns & 12.5 MHz & 1.20\% \\
		20000 & 48 ns & 10.4 MHz & 1.08\% \\
		15000 & 60 ns & 8.3 MHz & 1.00\% \\
		10000 & 96 ns & 5.2 MHz & 0.68\% \\
		\bottomrule
	\end{tabular}
\end{table}
\begin{figure*}[!t]
\centering
\includegraphics[width=0.98\textwidth]{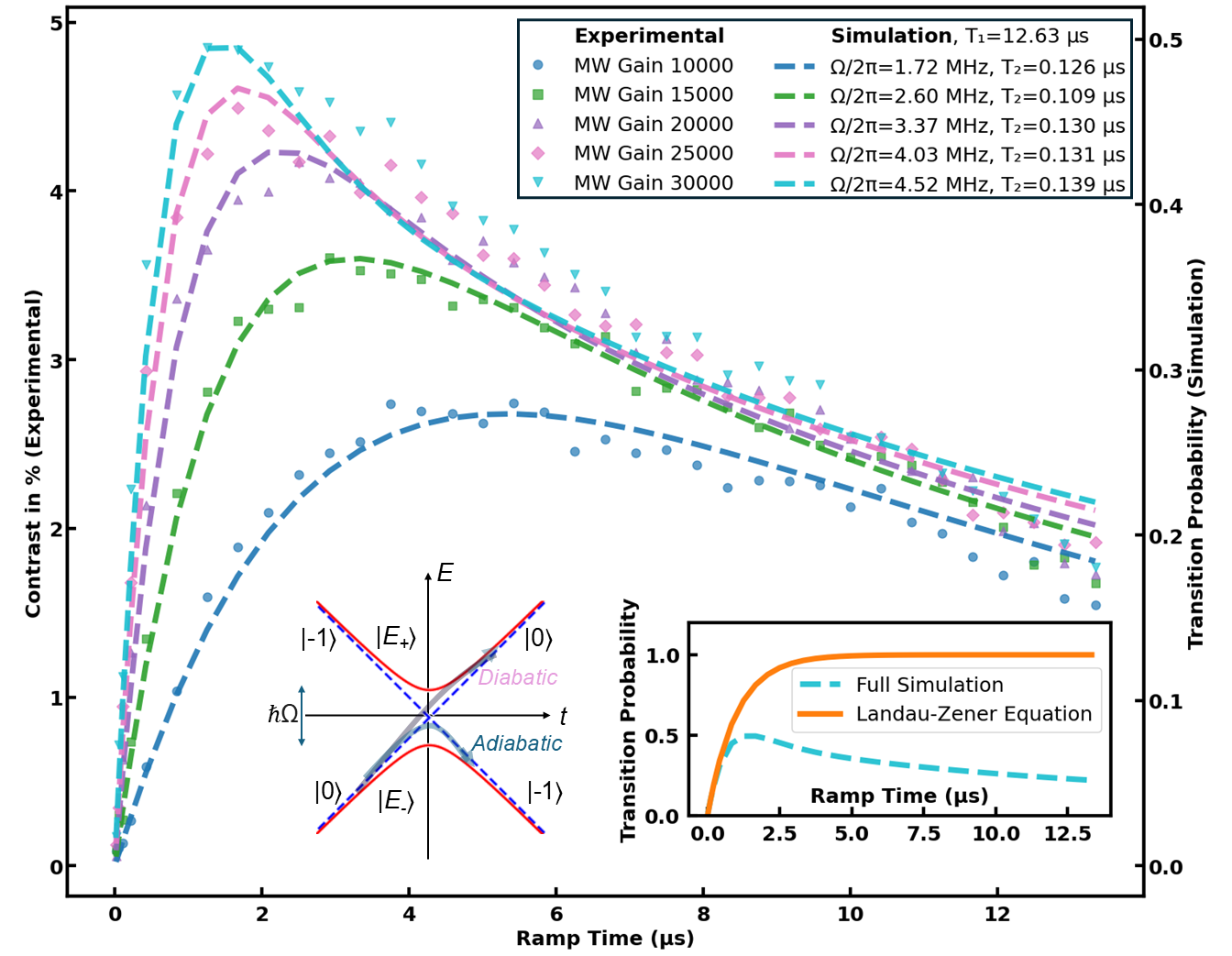}
\caption{Microwave gain and ramp time dependence of frequency-ramped spin inversion in V$_\text{B}^-$  h$^{10}$B$^{15}$N. Experimental contrast (measured at various MW power; left axis) demonstrates optimal ramp time for maximum contrast enhancement using frequency ramp of $\Delta f = 200$ MHz centered at 3160 MHz. Simulations via QuTiP Lindblad master equation (dashed lines, right axis) with optimized parameters (coupling $\Omega/2\pi$ and decoherence times T$_1$, T$_2$ for each gain level) fitted to experimental peak-normalized transition probabilities. Left inset: Schematic illustration of Landau-Zener transitions showing adiabatic versus diabatic pathways at an avoided crossing. Right inset: Comparison of full quantum simulation (dashed cyan) versus analytical Landau-Zener prediction (solid orange) for MW gain 30000.  All measurements performed at collection time = 1 $\mu$s; simulations share T$_1$ = 12.63 $\mu$s across power levels.}
\label{fig:figure3}
\end{figure*}
\subsection*{Analysis of Single LZ Frequency Ramp}
Figure \ref{fig:figure3} presents the comprehensive characterization of spin inversion from a single LZ frequency ramp as a function of both microwave gain and ramp duration, employing the optimal parameters identified from Fig.~\ref{fig:figure2}: a frequency ramp of $\Delta f = 200$ MHz centered at 3160 MHz. This center frequency corresponds to the midpoint of the $|0\rangle \to |-1\rangle$ ODMR transition that is shifted to lower frequency under applied magnetic field, as illustrated in Fig.~1(a). The experimental data shows that higher microwave gain leads to both increased peak contrast and faster optimal ramp times. Specifically, at MW gain 10000, we observe a peak contrast of 2.74\% at a ramp time of 5.42 $\mu$s, and this contrast increases progressively with power, reaching 4.85\% at 1.25 $\mu$s ramp time for gain 30000. Compared with the resonant $\pi$-pulse contrast values in Table~\ref{tab:your_label}, the frequency-ramped approach provides an approximately 4-fold enhancement of inversion when deployed. The table also provides a measure of microwave magnetic field at the sample (proportional to Rabi coupling $\Omega_{\text{nominal}}$) corresponding to the MW gain.

The power-dependent reduction of optimal ramp times correlates directly with the increase in effective microwave coupling $\Omega$. The system dynamics can be understood through the dressed-state picture of Landau-Zener transitions: at the avoided crossing arises as the bare spin states $|0\rangle$ and $|-1\rangle$ are hybridized by the microwave coupling into dressed eigenstates $|E_-\rangle = \cos(\theta/2)|0\rangle - \sin(\theta/2)|-1\rangle$ and $|E_+\rangle = \sin(\theta/2)|0\rangle + \cos(\theta/2)|-1\rangle$, where the mixing angle $\theta$ is given by $\tan\theta = \Omega/\delta$ with $\delta$ representing the instantaneous detuning from resonance. These dressed states (shown in the left inset of Fig.~\ref{fig:figure3}) are separated by the energy gap proportional to $\Omega$. Stronger microwave drive increases $\Omega$, enlarging this anticrossing gap. For fixed adiabaticity, defined by the ratio $\Omega^2/\alpha$ where $\alpha = 2\pi\Delta f/T$ is the sweep rate, a larger $\Omega$ permits proportionally larger $\alpha$ (faster sweeps) while maintaining the same transition probability. Since our experiments use a fixed frequency span $\Delta f = 200$ MHz, the sweep rate scales as $\alpha \propto 1/T$, enabling shorter optimal ramp times at higher gain while preserving adiabatic character.

While this idealized Landau-Zener picture explains the qualitative trends, the V$_\text{B}^-$ center exhibits significant spin relaxation on timescales comparable to the experimental ramp durations, necessitating a full open-system dynamics treatment. We therefore fit all five datasets jointly using Lindblad master equation simulations implemented in QuTiP, minimizing the sum of squared errors between normalized experimental contrast and simulated transition probabilities. We enforce a shared $T_1$ across all power levels while allowing $\Omega$ and $T_2$ to vary with microwave gain. The optimization yields $T_1 = 12.63$ $\mu$s, slightly below but close to the typically reported room-temperature range of 15-18 $\mu$s for V$_\text{B}^-$ centers \cite{gong2024isotope,gottscholl2020initialization}. With relaxation occurring on this timescale, the shorter optimal ramp times associated with higher microwave gain result in reduced population loss and thus higher peak contrast. The extracted dephasing times span from $T_2 = 109$ ns to $T_2 = 126$ ns across different power levels with no systematic trend. For context, previous spin echo measurements on hBN reported $T_2 = 92$ ns for h$^{11}$BN and $T_2 = 115$ ns for h$^{10}$BN \cite{haykal2022decoherence}, with dual isotopic purification (h$^{10}$B$^{15}$N) yielding spin echo times up to 186 ns \cite{gong2024isotope}. Our values, obtained from frequency-ramped inversion dynamics and reflecting effective dephasing during the sweep, are consistent with the expected enhancement from isotopic purification. The right inset of Fig.~\ref{fig:figure3} compares our full quantum simulation with the analytical Landau-Zener formula for MW gain 30000, revealing that at longer ramp times the analytical expression, which assumes negligible relaxation, fails to capture the observed peak and decay. 

The quantum simulations show a discrepancy between fitted effective Rabi frequencies and those estimated from resonant $\pi$-pulse measurements (Table~\ref{tab:your_label}). For instance, at gain 25000, the resonant $\pi$-pulse duration of 40 ns suggests a nominal Rabi frequency of $\Omega_{nominal}/2\pi = 1/(2\times40\,\text{ns}) \approx 12.5$ MHz, whereas our simulations extract $\Omega/2\pi = 4.03$ MHz, which is approximately one-third of the pi-pulse based estimate. This reduction, persisting across all power levels with fitted values ranging from 1.72 MHz to 4.52 MHz, arises from the complex hyperfine structure of the V$_\text{B}^-$ center in hBN. The broad ODMR spectrum in Fig.~\ref{fig:figure2}(a) shows four distinct hyperfine transitions spanning approximately 200 MHz. During resonant $\pi$-pulse experiments, the narrow-band microwave excitation addresses primarily a single hyperfine transition, yielding the true on-resonance Rabi frequency for that specific transition. In contrast, the frequency-ramped approach simultaneously sweeps through all four hyperfine components, and the fitted $\Omega$ represents an \textit{effective} parameter that averages over this multi-transition manifold, accounting for varying coupling strengths and detunings across the hyperfine spectrum. This is consistent with the observation that the frequency ramp provides more uniform inversion across the entire excitation spectrum as demonstrated by the robust contrast in Fig.~\ref{fig:figure2}(d) for varied center frequency of the ramp.

\begin{figure}[!htbp]
\centering
\includegraphics[width=\columnwidth]{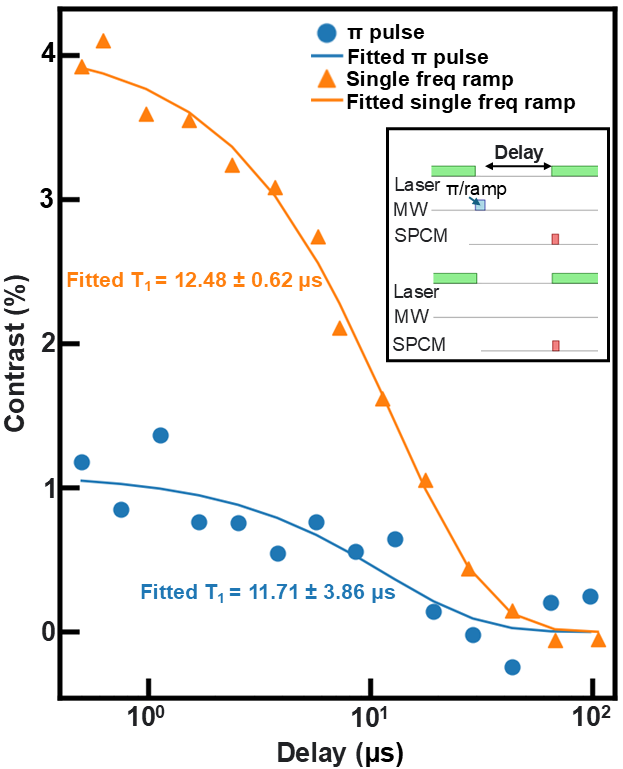}
\caption{Comparison of single frequency-ramped microwave pulse versus resonant-frequency microwave pulse spin manipulation for spin-lattice relaxation measurements in  V$_\text{B}^-$  h$^{10}$B$^{15}$N. Contrast decay measurements (circles: single frequency-ramped pulse with $\Delta f = 200$ MHz centered at 3160 MHz; triangles: resonant-frequency $\pi$-pulse at 3191 MHz) are fitted with exponentials (solid lines) to extract longitudinal relaxation time T$_1$. The frequency-ramped approach demonstrates four-fold contrast enhancement intially and overall superior signal-to-noise ratio. Data acquired at MW gain 25000 and collection time = 1 $\mu$s, with the inset showing the pulsing sequence for microwave on and off respectively, for obtaining the contrast.}
\label{fig:figure4}
\end{figure}

Figure \ref{fig:figure4} demonstrates the practical advantage of frequency-ramped spin manipulation during a longitudinal relaxation (T$_1$) measurements in V$_\text{B}^-$ h$^{10}$B$^{15}$N. We compare contrast decay curves obtained using two inversion protocols: a single frequency-ramped pulse with duration T = 1.67 $\mu$s, and $\Delta f = 200$ MHz centered at 3160 MHz (circles) versus a resonant-frequency $\pi$-pulse of duration of 40 ns at 3191 MHz (triangles), both at MW gain 25000. The data are fitted with single-exponential decays, yielding T$_1 = 12.48 \pm 0.62$ $\mu$s for the frequency-ramped approach and T$_1 = 11.71 \pm 3.86$ $\mu$s for the resonant $\pi$-pulse method, where the uncertainties represent one standard deviation from the covariance matrix of the nonlinear least-squares fit. These values are consistent within their uncertainties and confirm the T$_1 = 12.63$ $\mu$s extracted from the joint Lindblad fit in Fig.~\ref{fig:figure3}. The frequency-ramped approach achieves approximately a six-fold reduction in uncertainty, reflecting its superior signal-to-noise ratio and enabling more precise relaxometry measurements.

Thus the key distinction between the two approaches lies not in the extracted relaxation time but in the measurement quality. The frequency-ramped inversion exhibits an initial contrast of approximately 4\%, representing a four-fold enhancement over the resonant $\pi$-pulse contrast of around 1\% for initial delays. Visually, the T$_1$ relaxation data obtained using frequency-ramped inversion exhibit markedly reduced scatter compared to those obtained using resonant pi-pulse inversion, bolstered by the uncertainty differences between the obtained T$_1$ fits. Since both datasets were acquired under otherwise identical conditions, i.e. same microwave gain, same number of measurement repetitions per delay point, and same delay and readout time, this disparity in noise performance directly reflects the superiority of the frequency-ramped protocol. The measurement time for the whole T$_1$ measurement is dominated by the variable delay time after inversion rather than by the brief inversion pulse itself (whether 1.67 $\mu$s for the ramp or 40 ns for the $\pi$-pulse), ensuring fair comparison of signal quality per unit time. In precision measurements, SNR improves only as the square root of averaging time when reducing statistical noise through repeated acquisitions. The frequency-ramped approach delivers approximately four-fold SNR improvement over resonant pulses. To achieve equivalent measurement precision using resonant $\pi$-pulses would therefore require $16$ times more averaging, corresponding to a sixteen-fold increase in total acquisition time. 
\subsection*{Multiple LZ Frequency Ramps}
\begin{figure}[!htbp]
\centering
\includegraphics[width=\columnwidth]{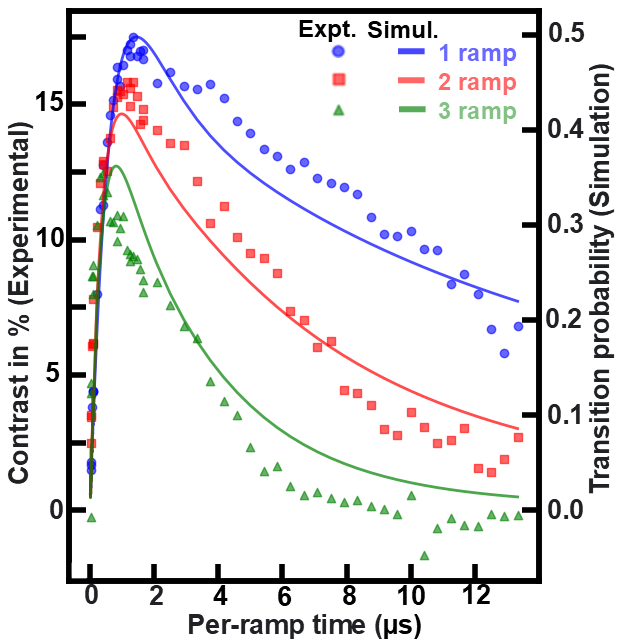}
\caption{Multi-sweep frequency-ramp induced $|0\rangle \rightarrow |-1\rangle$ spin inversion in V$_\text{B}^-$ defects in h$^{10}$B$^{15}$N. Contrast measurements (experimental data: circles for single sweep, squares for double consecutive sweeps, triangles for triple consecutive sweeps; left axis) showing leftward shift in optimal ramp time and reduced peak contrast with increasing number of sweeps. QuTiP Lindblad simulations of transition probability (solid lines, right axis) for consecutive forward frequency ramps with $\Delta f = 200$ MHz centered at 3160 MHz, coupling $\Omega/2\pi = 4.52$ MHz, T$_1$ = 12.6 $\mu$s, and T$_2$ = 0.139 $\mu$s, corroborating the experimental trends. Data acquired at MW gain 30000 and collection time 20 ns.}
\label{fig:figure5}
\end{figure}

Figure \ref{fig:figure5} extends our analysis to multi-sweep frequency-ramps for spin inversion in V$_\text{B}^-$ h$^{10}$B$^{15}$N. We reduce the photon collection time to 20 ns to minimize the effect of optical pumping during collection. While this yields substantially enhanced peak contrast of 17.5\% compared to the 1 $\mu$s collection window employed in Fig.~\ref{fig:figure3} for the single frequency-ramp induced inversion (circles), the enhancement is retained at around four-fold over resonant $\pi$-pulse inversion, which exhibits contrast of around 4\% under these conditions as shown in Supplementary Information (Fig. S2 (b)). The consecutive ramps presented here employ frequency sweeps in the same direction. We observe no significant dependence of the inversion efficiency on whether consecutive ramps sweep in the same or opposite directions (data not shown).

Using the optimized parameters extracted at MW gain 30000 ($\Omega/2\pi = 4.52$ MHz, T$_1 = 12.6$ $\mu$s, T$_2 = 0.139$ $\mu$s), we perform QuTiP Lindblad simulations (solid lines, right axis) for consecutive forward frequency ramps: single sweep (same simulation as plotted in Fig.~\ref{fig:figure3}), double consecutive sweeps, and triple consecutive sweeps, each with $\Delta f = 200$ MHz centered at 3160 MHz. The experimental data (circles for single sweep, squares for double sweep, triangles for triple sweep; left axis) reveal a systematic leftward shift in the optimal ramp time per sweep and a progressive reduction in peak contrast as the number of consecutive sweeps increases. The quantum simulations accurately reproduce these trends, confirming that the observations arise from the interplay of adiabatic passage dynamics and decoherence. Some quantitative differences arise between the predictions of our LZ model and experimental results for multiple sweeps due to non idealities during experiment such as non-instantaneous frequency change at the beginning of next consecutive ramp.

Unlike Landau-Zener interferometry schemes, where multiple passages through an avoided crossing can generate population oscillations as a function of sweep parameters, our system exhibits no such interference fringes. For a double-sweep protocol in the coherent limit, the transition probability follows \cite{shevchenko2010review}
\begin{equation}
	P_2(T) = 4P_{\text{LZ}}(T)(1-P_{\text{LZ}}(T))\sin^2\left(\frac{\Phi(T)}{2}\right),
\end{equation}
where $\Phi(T)$ is the accumulated Stückelberg phase between crossings. However, including significant dephasing, the double-sweep inversion probability can be approximated as $2P_{\text{LZ}}(T)\left(1-P_{\text{LZ}}(T)\right)\left[1-e^{-T/T_2}\cos\Phi(T)\right]$, derived by applying the identity $\sin^2(\Phi/2) = \tfrac{1}{2}(1-\cos\Phi)$ and introducing the decoherence envelope $e^{-T/T_2}$ during the time between crossings. In the strong-dephasing limit $T_2 \ll T$, this simplifies to
\begin{equation}
	P_2(T) \approx 2P_{\mathrm{LZ}}(T)\left(1-P_{\mathrm{LZ}}(T)\right),
\end{equation}
which reaches a maximum at $P_{\mathrm{LZ}} = 0.5$.

In the absence of $T_1$ relaxation, the single-pass Landau-Zener probability $P_{\mathrm{LZ}}(T)$ increases monotonically toward its asymptotic limit without exhibiting a peak, whereas the double-sweep function $2P_{\mathrm{LZ}}(1-P_{\mathrm{LZ}})$ has a maximum at $P_{\mathrm{LZ}} = 0.5$. Crucially, for small $P_{\mathrm{LZ}}$ values, the double-sweep probability $P_2 = 2P_{\mathrm{LZ}}(1-P_{\mathrm{LZ}})$ rises faster than the single-pass $P_{\mathrm{LZ}}$ itself. This faster rise means that $P_2$ reaches significant values at shorter sweep times $T$, where $T_1$ decay ($\propto e^{-T/T_1}$) has not yet substantially reduced the signal. Consequently, although the intrinsic maximum of $P_2$ is lower than the asymptotic limit of $P_{\mathrm{LZ}}$, the $T_1$-suppressed peak of $P_2$ (occurring at shorter $T$ with less decay) becomes closer in magnitude to the $T_1$-suppressed value of $P_{\mathrm{LZ}}$ at comparable times than would be expected in the absence of relaxation. This naive picture is consistent with the experimentally observed characteristics of the multi-sweep protocol: the peak appears at shorter sweep times and exhibits reduced height compared to single-sweep inversion, but the enhancement is partially preserved due to reduced $T_1$ decay at these shorter times.

For extended collection time, we surprisingly observe different optical pumping dependent decay for multiple LZ swept inversion, which is discussed more in supplementary information.

\section*{Conclusion and Outlook}\label{sec:conclusion}

In conclusion, we demonstrate frequency-ramped microwave pulses as a powerful and robust approach for enhanced spin manipulation of V$_{\text{B}}^{-}$ defects in isotopically enriched h$^{10}$B$^{15}$N. Using Landau-Zener transitions implemented via optimized chirped waveforms generated from FPGA, we achieve nearly a fourfold increase in population transfer and a corresponding sixteen-fold reduction in measurement time, while remaining robust to detuning and frequency drift. Numerical simulations show that the sweep dynamics are well captured by an effective two-level Landau-Zener model, despite the complex hyperfine structure. While increased microwave power primarily improves absolute population transfer, lower-temperature operation is expected to further enhance performance through extended $T_1$ times and improved adiabaticity.

Looking forward, this technique can be readily integrated into existing quantum sensing protocols using V$_{\text{B}}^{-}$ centers, is compatible with photonic cavity enhancements\cite{xu2023enhanced}, and is particularly well suited for low-signal or high noise environments, such as characterizing photosensitive samples. The FPGA-based implementation framework provides a flexible platform for exploring more sophisticated pulse sequences in future work. More broadly, frequency-ramped spin manipulation provides a general strategy for solid-state quantum systems with inhomogeneously broadened transitions, offering a versatile route to improved quantum sensing across diverse material platforms.

\section*{Methods}\label{sec:methods}

\subsection*{Sample Preparation}

Isotopically enriched h$^{10}$B$^{15}$N crystal flakes were grown by the atmospheric pressure high temperature (APHT) method, previously described in detail \cite{Liugrowth2018, Janzen2003}. Briefly, the process starts by mixing high purity 98\% enriched boron-10, with nickel and chromium with mass ratios of 3.72:48.14:48.14, respectively. The mixture is then heated at 200$^{\circ}$C/h under 97\% enriched nitrogen-15 and hydrogen gas at pressures of 787 and 60 torr respectively to 1550$^{\circ}$C, to produce a homogeneous molten solution. After 24 hours, the solution is slowly cooled at 1$^{\circ}$C/h, to 1500$^{\circ}$C, then at 50$^{\circ}$C/h to 1350$^{\circ}$C, and 100$^{\circ}$C/h to room temperature. The h$^{10}$B$^{15}$N solubility is decreased as the temperature is reduced, causing crystals to precipitate on the surface of the metal. The h$^{10}$B$^{15}$N flakes were exfoliated from the metal with thermal release tape.

The thermal release tape is further used to exfoliate h$^{10}$B$^{15}$N flakes on silicon substrate. Then they were bombarded with He$^{+}$ ion of energy 3~keV and $10^{15}$~cm$^{-2}$ fluence at \textit{CuttingEdge Ions}, a commercial facility.

The irradiated hBN flakes were subsequently transferred to a gold waveguide via a polycarbonate (PC) stamp attached to a polydimethylsiloxane (PDMS) support. For the transfer, the substrate was first heated to 160~$^{\circ}$C and subsequently cooled to 50~$^{\circ}$C before the PC stamp with adhered hBN flakes was lifted off. The stamp was positioned on the gold substrate and heated to 190~$^{\circ}$C, causing the PC film to melt, after which the PDMS support was detached. Finally, the substrate underwent sequential cleaning with chloroform, acetone, and isopropanol to achieve clean hBN flake deposition.

The gold on sapphire waveguide was produced through lithographic patterning utilizing an MLA 150 Maskless Aligner, with subsequent gold deposition performed via metal evaporation. A printed circuit board (PCB) for electrical connection to the waveguide was manufactured using an LPKF Protolaser U4.
\subsection*{Quantum Simulation Methodology}

To quantitatively capture the observed inversion vs ramp time in the experimental data, we employ full quantum dynamics simulations using the Lindblad master equation approach implemented via QuTiP \cite{johansson2013qutip}. The system is modeled as a driven two-level system with time-dependent Hamiltonian $H(t)/\hbar = -\frac{\Omega}{2}\sigma_x + \frac{A(t)}{2}\sigma_z$, where $\Omega$ represents the effective Rabi frequency and $A(t) = \frac{\Delta f \times 2\pi}{T} \cdot t$ describes the linear frequency sweep with total duration $T$ and frequency span $\Delta f$. The open system dynamics incorporate both energy relaxation (characterized by $T_1$) and pure dephasing (characterized by $T_2$) through Lindblad collapse operators with rates $\gamma_1 = 1/T_1$ and $\gamma_2 = 1/T_2 - 1/(2T_1)$. At zero temperature, the collapse operators are $C_1 = \sqrt{\gamma_1}\sigma_-$ (relaxation) and $C_3 = \sqrt{\gamma_2}\sigma_z$ (pure dephasing), where $\sigma_-$ is the lowering operator. Although our experiments are performed at room temperature, we set the thermal occupation number $n_{th} = 0$ in the simulations. This approximation is valid because the fitted $T_1$ and $T_2$ values are extracted directly from room-temperature experimental data and thus already incorporate all phonon bath effects at the measurement temperature. The thermal occupation parameter would be necessary only if modeling temperature-dependent dynamics or phonon-induced upward transitions explicitly, but since our decoherence parameters are temperature-specific phenomenological constants, $n_{th}$ can be neglected. Numerical integration employs adaptive step-size methods with 500 linearly spaced time points and a maximum of 10,000 internal steps to ensure convergence.

We perform joint multi-dataset parameter optimization across all five gain settings, enforcing a shared $T_1$ constraint while allowing gain-dependent $\Omega$ and $T_2$ values. The objective function minimizes the sum of squared errors between normalized experimental data and simulated probabilities across all datasets. Reference-based normalization preserves relative amplitude relationships between datasets: all experimental data and simulations are normalized by the maximum value of the highest-gain (30000) dataset, maintaining physically meaningful amplitude ratios across different power levels. A custom coordinate descent algorithm with bounded scalar minimization (Brent's method) efficiently navigates the high-dimensional parameter space (11 parameters: 1 shared $\gamma_1$, 5 per-dataset $\Omega$ values, and 5 per-dataset $\gamma_2$ values), converging to physically consistent solutions that simultaneously capture temporal dynamics and inter-dataset amplitude variations.

\backmatter

\bmhead{Data availability}

Data supporting the principal findings are available in the paper and Supplementary Information file. All raw datasets can be obtained from the corresponding authors on reasonable request.

\bmhead{Acknowledgements}

 This work was funded, in part, by the Laboratory Directed Research and Development Program and performed, in part, at the Center for Integrated Nanotechnologies, an Office of Science User Facility operated for the U.S. Department of Energy (DOE) Office of Science. Sandia National Laboratories is a multi-mission laboratory managed and operated by National Technology and Engineering Solutions of Sandia, LLC, a wholly owned subsidiary of Honeywell International, Inc., for the DOE’s National Nuclear Security Administration under contract DE-NA0003525. J.E. and T.P. acknowledge the support from the National Science Foundation, award number 2413808, for hBN crystal growth. We also thank U.S. Department of Energy, Office of Science, National Quantum Information Science Research Centers, Quantum Science Center for experimental work at Purdue University.
 Any subjective views or opinions that might be expressed in the paper do not necessarily represent the views of the U.S. Department of Energy or the United States Government. 

\bmhead{Author contributions}

M.A.S. designed samples and experiments, integrated frequency-ramped pulses into pulse sequences, performed simulations and optimization for analysis, and drafted the manuscript. T.D. developed FPGA code for frequency ramp generation. L.B. established optical, magnetic, and microwave delivery setups. T.P. and J.E. synthesized h$^{10}$B$^{15}$N via crystal growth. J.H. contributed to Landau-Zener analysis. P.A.B. contributed to enhancement and noise analysis and manuscript proofreading. Y.P.C. provided supervision and guidance on analyzing Landau-Zener physics. A.M. conceived and supervised the project, designed experiments, and helped establish the simulation framework and analysis of the Landau-Zener dynamics.

\bmhead{Competing interests}

The authors declare no competing interests.

\bmhead{Supplementary information}
Supplementary Information is available for this paper.

\bibliography{sn-bibliography}


\end{document}


\maketitle

\setcounter{figure}{0}
\section*{Optical and microwave signal generation and processing}
\begin{figure}[H]
    \centering
    \renewcommand{\thefigure}{S\arabic{figure}}
    \includegraphics[width=0.85\textwidth]{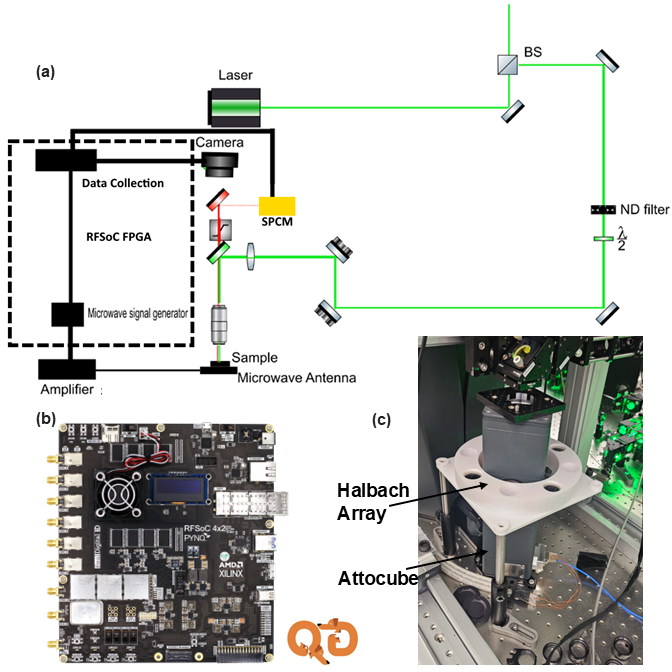}
    \caption{Experimental setup for phase-controlled frequency-ramped microwave excitation of V$_\text{B}^-$ defects in hBN. (a) Schematic diagram of the experimental apparatus, including optical excitation and collection pathways, microwave generation and delivery systems, and signal processing components. (b) RFSoC 4$\times$2 FPGA controlled via the Qick-Dawg package for phase-controlled frequency ramping, precise timing sequences, and real-time data acquisition. (c) A Halbach array generates the external magnetic field, and the device is mounted in an attocube cryostat system under vacuum for sample positioning and microwave delivery.}
    \label{fig:figureS1}
\end{figure}

Our experimental apparatus employs a 532~nm green laser to illuminate the sample, with emitted red light collected through a longpass filter and directed to a single photon counting module (SPCM) for detection of V$_\text{B}^-$ centers in hBN (Fig.~\ref{fig:figureS1}(a)). The microwave control system utilizes an FPGA-based RFSoC 4x2 platform (Fig.~\ref{fig:figureS1}(b))) integrated with the Quantum Instrument Control Kit - Defect Arbitrary Waveform Generator (QICK-DAWG)~\cite{riendeau2023quantum}---an open-source control framework originally developed at Sandia National Laboratories for nitrogen-vacancy center manipulation---to synthesize and deliver microwave signals, which are subsequently amplified using a dedicated microwave amplifier. We adapted QICK-DAWG for investigating V$_\text{B}^-$ defects in hBN, incorporating support for frequency-ramped pulse sequences. The FPGA additionally serves as the data acquisition and processing unit. To generate a uniform magnetic field perpendicular to the hBN plane, we fabricated, using Ender 3 S1 plus 3D printer, a plastic housing into which four permanent magnets were installed in a Halbach Array configuration along the circumference of a circle, with the hBN sample positioned at its center. The sample is housed within an evacuated attocube chamber to minimize environmental perturbations (Fig.~\ref{fig:figureS1}(c)).

\newpage

\section*{Observation of slower decay under optical pumping of inverted state resulting from multiple frequency ramp}

\begin{figure}[H]
	\centering
	\renewcommand{\thefigure}{S\arabic{figure}}
	\includegraphics[width=0.95\textwidth]{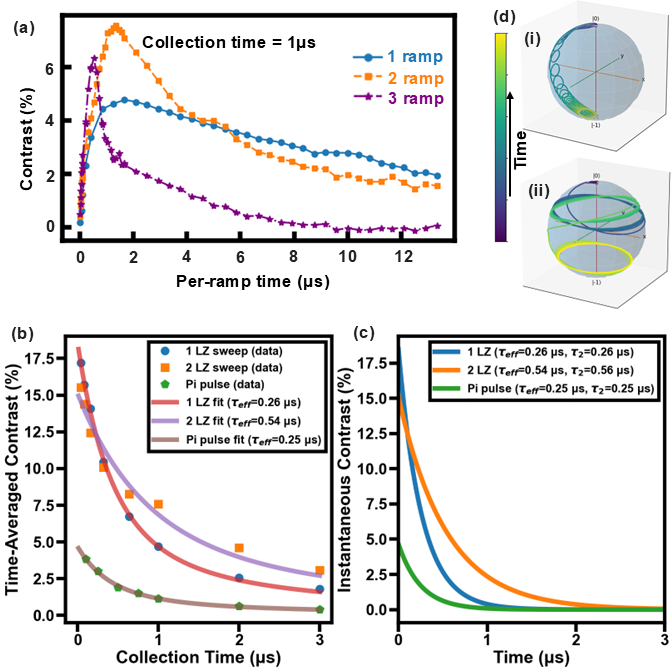}
	\caption{Collection time-dependent contrast decay for consecutive Landau-Zener sweeps. (a) Contrast from inversion as a function of per-sweep time for single (blue circles), double (orange squares), and triple (purple stars) consecutive Landau-Zener sweeps, measured with MW gain 30000 and 1~$\mu$s photon collection time. (b) Time-averaged contrast decay versus total collection time for three distinct inversion protocols: $\pi$-pulse (green pentagons), single Landau-Zener sweep (blue circles), and double consecutive Landau-Zener sweeps (orange squares). Experimental data (markers) are fitted to an analytical model (solid lines) that accounts for simultaneous T$_1$ relaxation ($\tau_1 = 12.63~\mu$s) and protocol-dependent decay characterized by effective decay times $\tau_\text{eff}$ and optical pumping induced decay $\tau_2$. (c) Reconstructed instantaneous population dynamics for the same three protocols ($\pi$-pulse, single sweep, double sweep) derived from the fitted effective decay times, showing slower optical pumping induced decay for multiple frequency sweep (d) Bloch sphere illustrations showing (i) slower single sweep trajectory and (ii) faster double sweep trajectory necessary for peak inversions, with lighter shading indicating temporal progression. The final state from double sweep from peak inversion tend to exhibit more equitorial nature compared to that of the single sweep.}
	\label{fig:sweep_optimization}
	
	\label{fig:figureS2}
\end{figure}

We observe a dependence of the apparent optical pumping decay rate on the number of consecutive Landau-Zener (LZ) sweeps employed for spin state inversion. Figure~\ref{fig:figureS2} demonstrates that multiple-sweep protocols exhibit substantially slower decay under continuous optical excitation compared to single-sweep protocols. Figure~\ref{fig:figureS2}a presents the contrast from inversion as a function of per-sweep time for single (blue circles), double (orange squares), and triple (purple stars) consecutive LZ sweeps, measured with microwave gain 30000 and 1~$\mu$s photon collection time. Surprisingly, the peak contrasts for multiple LZ ramps under these experimental conditions are greater than that of the single LZ ramp.

To quantify these dynamics, we analyze the time-averaged contrast decay versus total collection time for three distinct inversion protocols: $\pi$-pulse (green pentagons), single LZ sweep (blue circles), and double consecutive LZ sweeps (orange squares), as shown in Figure~\ref{fig:figureS2}b. The experimental data (markers) are fitted to an analytical model (solid lines) accounting for two parallel relaxation pathways:

\begin{equation}
	\frac{1}{\tau_{\text{eff}}} = \frac{1}{\tau_2} + \frac{1}{\tau_1}
\end{equation}

\noindent where $\tau_1 = 12.63~\mu$s is the spin-lattice (T$_1$) relaxation time, and $\tau_2$ represents the protocol-dependent optical pumping induced decay time constant. The time-averaged signal over a measurement window $[0, T]$ is given by:

\begin{equation}
	\langle S(T) \rangle = A \cdot \frac{\tau_{\text{eff}}}{T} \left( 1 - e^{-T/\tau_{\text{eff}}} \right)
\end{equation}

\noindent where $A$ denotes the initial amplitude. This model provides satisfactory fits to the experimental decay curves across all three protocols, yielding protocol-specific effective decay times $\tau_{\text{eff}}$ from which we extract the optical pumping time constants $\tau_2$.

Figure~\ref{fig:figureS2}c presents the reconstructed instantaneous population dynamics derived from the fitted effective decay times. While all three protocols achieve initial population inversion, the double-sweep protocol exhibits substantially slower optical pumping induced decay ($\tau_2 \approx 0.56~\mu$s) compared to the single-sweep protocol ($\tau_2 \approx 0.26~\mu$s)---a reduction in decay rate by approximately a factor of two. The single- and double-sweep protocol-specific $\tau_2$ values were extracted by fitting the time-averaged decay curves in panel (b) to obtain $\tau_{\text{eff}}$ for each protocol, then employing the known $\tau_1 = 12.63~\mu$s to calculate $\tau_2$ from the relation $1/\tau_{\text{eff}} = 1/\tau_2 + 1/\tau_1$. The triple-sweep protocol shows $\tau_2 \approx 0.60~\mu$s, only marginally slower than the double sweep; this value was obtained from earlier data showing contrast decay from ~12\% at 20~ns collection time (Fig.~5) to ~6\% at 1~$\mu$s collection time (Figure~\ref{fig:figureS2}a) . The $\pi$-pulse protocol exhibits similar optical pumping decay rate ($\tau_2 \approx 0.25~\mu$s) post microwave pulse as single frequency ramp microwave pulse.

Numerical simulations of the density matrices at peak inversion times indicate that the off-diagonal coherence term $|\rho_{01}|$ is approximately 1.6 times larger for double and triple sweep protocols compared to single sweep protocols. States characterized by these larger off-diagonal terms exhibit approximately 2 times slower apparent optical pumping rates. The saturation observed between double and triple sweep protocols suggests the coherence magnitude approaches a maximum value determined by the populations.

Figure~\ref{fig:figureS2}d illustrates the distinct sweep trajectories on the Bloch sphere. The final state from double sweep at peak inversion exhibits, besides a lower degree of spin  inversion compared to single sweep, a more equatorial character---indicating enhanced off-diagonal density matrix elements---compared to the predominantly polar (diagonal) character of the single-sweep final state.

The observed correlation between larger off-diagonal coherence $|\rho_{01}|$ and slower optical pumping decay is consistent with phenomena reported in other atomic systems, where off-diagonal elements have been shown to modify optical pumping dynamics through mechanisms such as coherent population trapping\cite{arimondo_coherent_1996,fleischhauer_electromagnetically_2005}. However, these systems differ significantly from V$_{\text{B}}^{-}$ centers in hBN in both electronic structure and excitation/relaxation pathways, rendering direct analogies potentially problematic. A rigorous theoretical treatment of coherence-dependent optical pumping in the V$_{\text{B}}^{-}$ system would require detailed modeling of the excited state manifold and metastable state pathways, which lies beyond the scope of this work.

As this phenomenon has not been rigorously modeled analytically, the claim for coherence-dependent optical pumping remains tentative, and the observed difference in decay rates could also arise from hidden timing errors in the pulse sequence implementation. The execution of multiple consecutive sweeps introduces substantial complexity in pulse programming and timing sequences. Systematic differences in photon collection windows between single- and multiple-sweep protocols, or accumulated timing offsets during long collection periods, could contribute to the apparent differences in decay rates.

If the slower decay rate for multiple frequency ramps is not an experimental artifact, multiple-sweep protocols, particularly double sweeps, may provide substantially improved signal-to-noise ratio for inversion measurements requiring extended collection times. This capability would prove especially valuable under challenging experimental conditions such as low laser power necessitated by photosensitive samples, low quantum efficiency detection systems, or samples exhibiting inherently weak photoluminescence. The saturation observed between double and triple sweeps indicates that protocols employing more than two consecutive sweeps yield diminishing returns, providing guidance for optimal experimental design.

\bibliography{sn-bibliography}